\documentclass[aps,showpacs,twocolumn,floatfix,superscriptaddress,reprint]{revtex4-1} 
\usepackage{graphicx} 
\usepackage{amssymb}

\newcommand \be {\begin{equation}}
\newcommand \ee {\end{equation}}
\newcommand \bea {\begin{eqnarray}}
\newcommand \eea {\end{eqnarray}}

\begin{document}

\title{Clustering and anisotropic correlated percolation in polar flocks} 

\author{Nikos Kyriakopoulos}
\affiliation{Department of Applied Physics, Aalto University,
  Helsinki, Finland}
\author{Hugues Chat\'e} 
\affiliation{Service de Physique de l'Etat Condens\'e, CEA, CNRS, Universit\'e Paris-Saclay,
CEA-Saclay, 91191 Gif-sur-Yvette, France}
\affiliation{Beijing Computational Science Research Center, Beijing 100094, China}
\author{Francesco Ginelli}
\affiliation{Department of Physics and Institute for Complex Systems and Mathematical Biology, King's College, University of Aberdeen, Aberdeen AB24 3UE, United Kingdom}


\begin{abstract}
We study clustering and percolation phenomena in the Vicsek model, 
taken here in its capacity of prototypical model for dry aligning active matter.
Our results show that the order-disorder transition is not related in any way to a percolation transition, contrary to some earlier claims.
We study geometric percolation in each of the phases at play, but we mostly focus on the ordered Toner-Tu phase,
where we find that the long-range correlations of density fluctuations 
give rise to a novel anisotropic percolation transition.
\end{abstract}

\maketitle 

\section{Introduction}

Active matter \cite{Ramaswamy2010} typically involves moving
``particles" (such as social animals \cite{Ballerini2008},
cells \cite{Hakim, Trepat}, biofilaments displaced by motor proteins \cite{Schaller2010}, phoretic
colloids \cite{Palacci}, etc.). Energy, either stored internally or gathered from the environment, is consumed
locally to produce mechanical work.
These systems display a wide range of collective phenomena that are not possible  
in equilibrium. In particular, Toner and Tu have shown that flocking systems
such as the celebrated Vicsek model \cite{Vicsek95},
where constant-speed particles locally align their velocities in the presence of noise, 
can show true long-range orientational order even in 2 dimensions, in a strongly fluctuating phase endowed by 
generic anisotropic long-range correlations \cite{TT1,Rama-GDF,
  Ginelli2008, CavagnaScaleFree, Giavazzi2017}. 
 
Active matter systems are also known to often show dense clusters that dynamically form, merge, shrink, and split.
This has been observed experimentally in situations as diverse as 
bacteria colonies \cite{Swinney2010}, acto-myosin motility assays
 \cite{Schaller2010, Schaller2010b,Frey-Science}, animal groups
\cite{PNAS}, and active colloidal particles \cite{Palacci2012}.
A wide, powerlaw-like, distribution of cluster sizes has been reported in certain cases
such as gliding myxobacteria \cite{Peruani2012}. Simple models of self-propelled rods interacting solely via steric exclusion,
put forward initially in the context of bacteria, have long been known to exhibit similarly broad distribution of cluster sizes
 \cite{Peruani2006,GOMPPER,GOMPPER2}, a situation sometimes referred to as
non-equilibrium clustering. 
In most of these systems, these clusters are believed to be the consequence of arrested --or micro-phase separation,
with size or mass distributions bounded by a finite, albeit sometimes very large, intrinsic cut-off
\cite{NARDINI,Shi2018}.

Clusters also appear in flocking models such as the Vicsek model, where they are naturally and unambiguously  defined
by making use of the finite-range of interactions.
Power-law distributions of cluster sizes have also been reported
 \cite{Huepe, Ginelli2008, Peruani2008}. Because these observations were mostly made
in the region of parameter space where the order-disorder transition takes place,  
some authors have conjectured that, in active systems exhibiting collective motion, this transition from disorder to
ordered collective motion could be somehow generically related to (or even 
mediated by) non-equilibrium clustering\cite{Peruani2013}. 
This claim, at face value, may appear rather surprising: indeed, in a noisy model such as the Vicsek model,
one expects that at large enough density, particles would always form a single, macroscopic, spanning cluster, 
irrespective of the degree of orientational order present. Conversely, at low enough densities, one has no chance to observe
a percolating cluster. It is thus natural to expect a percolation transition
\cite{Stauffer} separating these two regimes. 

Moreover, phase-separation has been recently shown to be at play in dry aligning active matter. 
It actually provides the best framework to understand the phase diagram of Vicsek-style models \cite{Tailleur2,ChateReview}, 
which contain 3 phases, with a disordered gas separated from an ordered liquid by a coexistence phase.
To the best of our knowledge, it remains unclear whether geometric percolation and the order-disorder transition
can interfere in any way in flocking models.

In this work, we come back to this issue, and study clustering phenomena 
in the Vicsek model, taken here as a prototypical model for dry aligning active matter.
Our results  show that the order-disorder transition is {\it not} related in any way to a percolation transition.
We study geometric percolation in each of the phases at play, but we mostly focus on the ordered Toner-Tu phase,
where we find that the long-range correlations of density fluctuations
give rise to a novel anisotropic percolation transition.

The remainder of this paper is organized as follows: 
in Section~\ref{sec:VM}, we summarize the phase diagram of the Vicsek model and recall some of its basic properties.
Sections~\ref{sec:clust},\ref{sec:FSS},\ref{sec:CSD} describe percolation and clustering in the Toner-Tu liquid phase, while we briefly 
examine the disordered and the coexistence phase in Section~\ref{sec:vcuts}. A discussion and some conclusions can be found in 
Section~\ref{sec:conclusion}.

\section{The Vicsek model for flocking and its phase diagram}
\label{sec:VM}

We consider the classic version of the Vicsek model \cite{Vicsek95} with metric 
interactions in two spatial dimensions. Particles are defined 
by an off-lattice position ${\bf r}_i$ and an orientation $\theta_i
\in [0, 2\pi]$, with $i=1,\ldots,N$.
The discrete-time evolution is synchronous:  orientations
and  positions  are updated at  integer
time steps according to the driven-overdamped dynamics
\be
\theta_i(t) = \mbox{Arg}\left[ \sum_{j=1}^N \mathcal{A}^t_{ij} {\bf v}_i(t)\right] +
\eta\, \xi_i(t)
\label{eq:theta}
\ee
\be
{\bf r}_i(t+1) = {\bf r}_i(t) + v_0 {\bf v}_i(t+1)\,,
\label{eq:r}
\ee
where ${\bf v}_i = \left(\cos (\theta_i), \sin (\theta_i)\right)$ is the unit
vector pointing in the direction $\theta_i$, $v_0$
is the speed of particles
and $\xi_i^t$ is a random angle drawn uniformly in $[-\pi,\pi]$
with delta correlations in space and time.
The alignment interaction is limited to a metric range with a
radius $r_0=1$ \cite{noteS}, and the symmetric and time-dependent interaction matrix
$\mathcal{A}^t$ codes for the presence of neighbors within this interaction range:
\begin{equation}
\mathcal{A}^t_{ij} = \left\{
\begin{array}{c c}
1 & \mbox{    if   } ||  {\bf r}_i(t) -  {\bf r}_j(t)|| \leq 1 \\
&\\
0 & \mbox{    if   } ||  {\bf r}_i(t) -  {\bf r}_j(t)|| > 1 \,.
\end{array}\right.
\end{equation}
Effectively, the sum in Eq. (\ref{eq:theta}) runs
over all particles in the unit radius disk
centered around particle $i$ ($i$ itself included).
The finite interaction radius $r_0$ allows for a natural and unambiguous definition of clusters: 
particles within distance $r_0$ of each other belong to the same cluster.
At any given time $t$ clusters are then determined 
as the connected components of the graph formed by the interaction matrix $\mathcal{A}^t_{ij}$.

We consider square domains of linear size $L$ with periodic boundary
conditions, corresponding to a global density $\rho=N/L^2$.
In the following we fix $v_0=0.5$ and consider the usual two main control parameters,
the global density $\rho$ and the noise amplitude $\eta$, the latter playing
a role akin to that of temperature in equilibrium systems. 

For maximum noise, $\eta=1$, particle orientations are completely
random and decorrelated, so that at each time step their spatial distribution
is equivalent to one drawn from a Poisson point process
\cite{Gabrielli}. As the noise is lowered, short range correlations initially
build up (both in orientation and position) and, as a threshold
$\eta_{\rm gas}$ is passed, the system eventually undergoes a spontaneous
symmetry breaking phase transition to long-ranged (polar) order,
easily characterized by the mean particle orientations order parameter,
${\bf V}(t) = \frac{1}{N}  \sum_i^N {\bf v}_i(t)$.

\begin{figure}[t!]
 \includegraphics[width=\columnwidth]{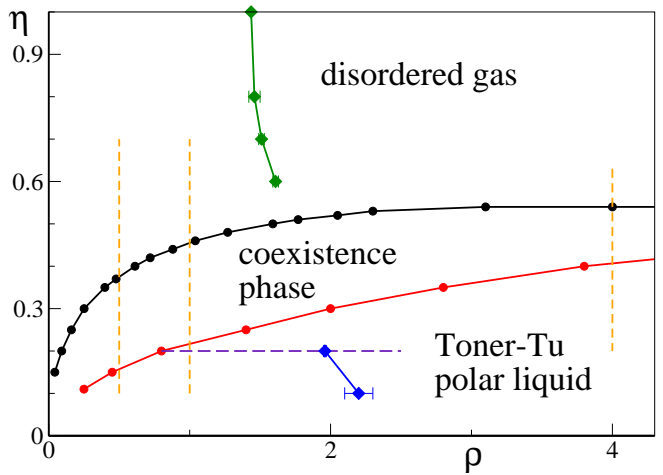} 
\vspace{0.01cm}
 \caption{Phase diagram of the Vicsek model for $v_0=0.5$ in the
   $(\rho, \eta)$ plane (from Ref.~\cite{Solon2015}). 
   The binodal line $\eta_{\rm gas}(\rho)$ separating the disordered
   gas from the coexistence phase made of traveling high-density high-order bands 
   is reported in black, while the red line marks the liquid binodal $\eta_{\rm liq}(\rho)$ separating 
   the coexistence region from the Toner-Tu polar liquid. 
   The green line links the diamonds locating the isotropic percolation threshold in the disordered gas phase.
   The 2 blue diamonds linked by the solid line show the asymptotic location of the percolation
   transition in the Toner-Tu liquid phase determined through finite size scaling
   (see Sec. \ref{sec:FSS}).
   The $\eta=0.2$ horizontal indigo dashed line illustrates the parameter line
   investigated in detail in Sections \ref{sec:clust}-\ref{sec:FSS}.
   The vertical orange lines mark the density values analyized in Section \ref{sec:vcuts}.}
\label{PD}
 \end{figure}

Active particles move
following the orientational degrees of freedom that they themselves
carry, linking local order and local density in a simple but highly
non-trivial way. As a result, the transition between
the fluctuating but {\it homogeneous} disordered and ordered phases is
not direct, like originally thought in analogy with magnetic systems such as the XY model,
but mediated by a coexistence phase where high-density 
ordered bands move in a low-density disordered background \cite{Gregoire2004,  Ginelli2008, Bertin2,Ihle}. 
Within the coexistence phase, increasing the global density and/or the system size, the number of traveling bands increases linearly
while the residual vapor density between them remains constant \cite{Solon2015}. We are thus in the 
presence of a phase separation scenario: the disordered gas (DG) is
separated from the ordered Toner-Tu polar liquid (PL) by a coexistence 
region with a quantized liquid fraction (the traveling bands are microphases).
The corresponding asymptotic phase diagram, following the numerical results of
\cite{Solon2015}, is reported in Fig.~\ref{PD}. 
One has thus two transitions, not one, marked by the two binodal lines
separating these different phases. They are non-decreasing
functions of density, $\eta_{\rm gas}=\eta_{\rm gas}(\rho)$,
$\eta_{\rm liq}=\eta_{\rm liq}(\rho)$, and in the limit of small densities one has $\eta_{\rm gas} \sim \sqrt{\rho}$ \cite{VPM}.
An inaccessible critical point is pushed towards infinite density \cite{Tailleur2}. 
The two transitions are continuous (in the infinite-size limit) but not critical. At finite size, they appear discontinuous
because of the large number of particle involved in nucleating a traveling band.

\section{Clustering and anisotropic percolation in the Toner-Tu liquid phase}
\label{sec:clust}

\begin{figure*}[t!]
 \includegraphics[width=0.9\textwidth]{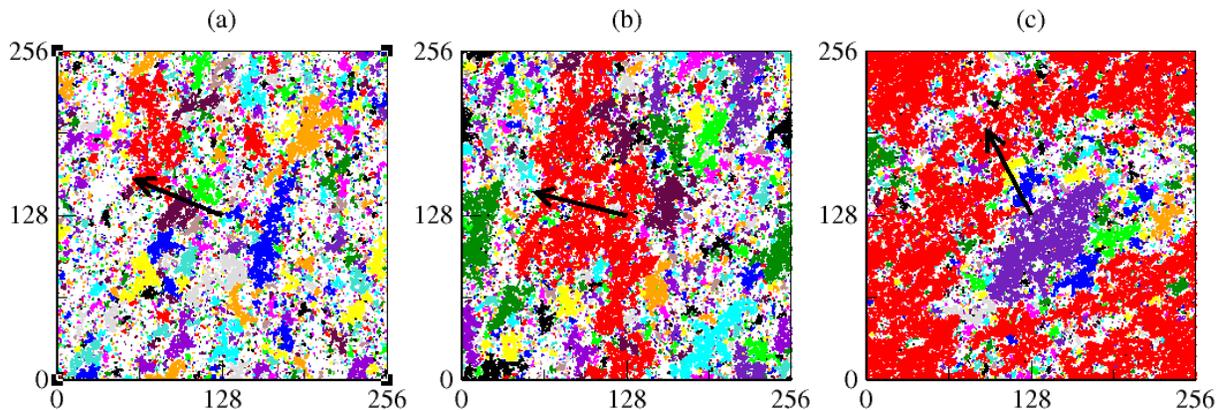} 
 \caption{Typical instantaneous snapshots in the Toner-Tu ordered phase at different densities
 ((a) $\rho=1$, (b) $\rho=1.5$, $\rho=1.9$). Other parameters $v_0=0.5$, $\eta=0.2$.
  Colors correspond to connected clusters of particles, with the largest cluster in red.
  Note that due to the large number of 
   different clusters (in the order of thousands in all three
   panels), each color is used for several distinct clusters, hopefully sufficiently apart from each other to avoid confusion.
The thick black arrow marks the instantaneous
direction of global order (i.e. the order parameter orientation)}
\label{fig1}
 \end{figure*}

In the following 3 sections, we focus our attention on the polarly-ordered Toner-Tu liquid phase.
We initially fix the noise amplitude to $\eta=0.2$, and study the
clustering behavior for different particle densities $\rho > 0.8$, i.e. below the liquid binodal $\eta_{\rm liq}$ in Fig.~\ref{PD}.
Three typical  snapshots, obtained in the stationary
regime for increasing global densities ($\rho=1, \rho=1.5$ and $\rho=1.9$) 
are shown in Fig.~\ref{fig1} for a system of linear size $L=256$.

At the lowest density value $\rho=1$, the largest clusters are clearly smaller than system size.
The largest of them contains less than $10\%$ of the total number of particles.
The transversal extension
(with respect to the current global order direction) of these largest clusters is much larger than the longitudinal one.
Increasing the density, clusters remain clearly anisotropic and some of them are spanning across the system
along the transversal direction. 
In the central panel of Fig.~\ref{fig1}, one sees a single spanning cluster, comprising less than one third of the
total number of particles.   
It is only at densities $\rho \gtrsim 1.7$ that the largest cluster starts spanning across all directions, 
i.e. both transversally and longitudinally. The largest cluster then contains a large majority of all particles, as
clearly visible in Fig.~\ref{fig1}c. 

This brief graphical inspection suggests that there might be two distinct percolation thresholds, defined by 
the fact that the largest cluster first spans the system in the direction transverse to global order, and then spans it 
in all directions.
This anisotropy between the transverse and longitudinal directions
is not surprising; indeed it is known that
the Toner-Tu phase displays anisotropic scaling laws \cite{TT1}. For instance, the 
 two-point correlation functions
of density and velocity fluctuations display generic anisotropic algebraic decay:
\begin{equation}
C({\bf r}) = \left| {\bf r}_\perp \right|^{2\chi} f(r_\|/\left| {\bf r}_\perp \right|^{\zeta}) \,,
\end{equation}
where $\|$ and $\perp$ indices respectively refer to directions longitudinal and transverse to the mean motion of the flock
and the exponents $\chi$ and $\zeta$ as well as the function $f$ are universal. 
A notable consequence of this fact is that, in two spatial dimensions, the particles' displacement
transversal to the mean velocity is superdiffusive \cite{TT2, Ginelli2008}, while it is simply
diffusive in the longitudinal direction (once substracted the mean motion). 

We now characterize the percolation transition and its anisotropy from a more quantitative
point of view. Individual clusters (labeled by $k$) can be quantified by 
their mass $s_k$, that is, the number of particles in the cluster, and by their
linear extension $\ell_k$, which we define as twice the in-cluster maximum distance between a cluster particle and 
the cluster center of mass\cite{note1}. 
The instantaneous maxima of these quantity are respectively
$s_M=\max_k (s_k)$ and $\ell_M=\max_k (\ell_k)$
where the cluster index $k$ runs over all clusters of a given configuration.

\begin{figure}[t!]
 \includegraphics[width=\columnwidth]{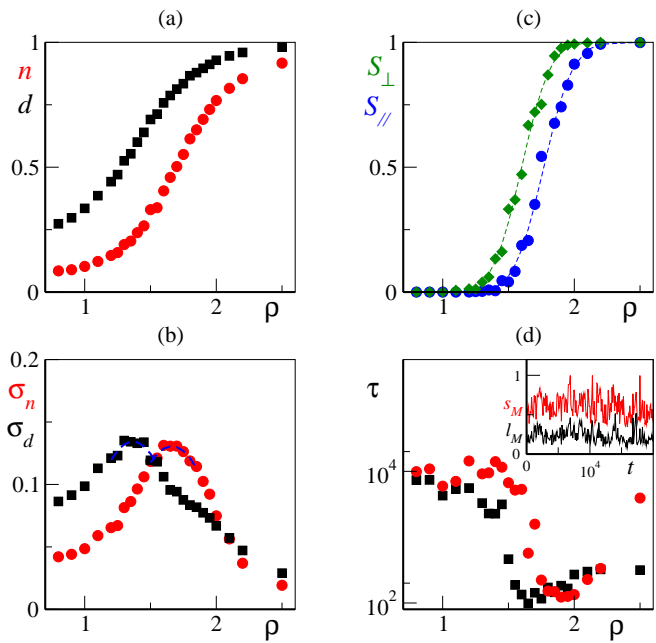} 
 \caption{Anisotropic percolation transition in the Toner-Tu ordered phase. 
 (a) Normalized mean
   largest cluster size $n$ (red dots) and normalized mean cluster maximum
   linear extension $d$ (black squares) as a function of the global density $\rho$. 
   (b) Corresponding standard deviations $\sigma_n$ (red dots)
   and $\sigma_d$ (black squares).
    The blue dashed lines show quadratic fits of the peak regions (see text). 
   (c) Transversal ($S_\perp$) and longitudinal
   ($S_\parallel$) spanning probabilities as a function of total
   density. The dashed lines show a fit based on the error function
   (see text). 
(d) Autocorrelation time $\tau$ of the timeseries of maximal cluster size $s_M$ (red dots)
   and maximal linear extension $\ell_M$ (black squares) as a function of global density.
   Inset: typical excerpts from these timeseries for $\rho=1.5$. Other parameters:
   $L=256$, $\eta=0.2$, and $v_0=0.5$. Configurations averages have been
   computed sampling every 100 timesteps a $T=10^6$ timeseries in the stationary regime.
 The standard error (see text) of the data shown in
   panels (a)-(c) is equal or smaller than the symbol size.}
\label{fig2}
 \end{figure}

Two order parameters are routinely employed in the literature \cite{Berthier2014} about 
isotropic percolation problems, the (normalized) mean largest cluster size $n$ and 
the mean cluster maximum linear extension $d$, where the average is taken over many different realizations
(e.g. sampling a long trajectory in the stationary state at regular time intervals). 
The definitions of $n$ and $d$ and the associated standard deviations $\sigma_n$ and $\sigma_d$ read:  
\begin{eqnarray}
n & \equiv \frac{\langle s_M \rangle}{N} \;,\;\;\; & \sigma_n \equiv \frac{\sqrt{\langle (s_M - \langle s_M \rangle)^2\rangle}}{N} 
\label{eq:n} \\
d & \equiv \frac{\langle \ell_M \rangle}{\sqrt{2}\,L} \;,\;\;\; & \sigma_d \equiv \frac{\sqrt{\langle (\ell_M - \langle \ell_M
    \rangle)^2\rangle}}{\sqrt{2}\,L} \,.
\label{eq:d}
\end{eqnarray}
In our anisotropic situation, $n$ and $d$ are expected to behave differently as the density is
increased, with $d$ rising earlier to order 1 values and $n$
following later. This is indeed observed in Fig.~\ref{fig2}a where the
two indicators are compared for a system of size $L=256$ and
$\eta=0.2$. 
The standard deviations $\sigma_n$ and $\sigma_d$ peak at two different density values (Fig.~\ref{fig2}b).

We also investigated directly the spanning probability $S$, i.e. 
 the probability that a spanning cluster does appear. 
 While, in the thermodynamic limit, $S(\rho)$ is a step function with
the jump exactly located at the phase transition, in finite systems
$S(\rho)$ is smoothed around the (finite size) percolation point \cite{Stauffer}. 
To take into account anisotropy, we consider both a transversal and a longitudinal spanning probability, 
$S_\perp$ and $S_\parallel$, defined respectively as the probability
that a cluster wraps around the $L \times L$ torus in the transversal 
or longitudinal (w.r.t. the order parameter ${\bf V}$) directions \cite{note2}. (We
discuss the accuracy of these measures in finite systems -- where
fluctuations lead to the diffusion of the instantaneous mean orientation of motion
${\bf V}(t)$ -- in the next section.) For the moment, we simply note that
the transversal spanning probability $S_\perp$ rises from zero
towards one earlier than the longitudinal probability
$S_\parallel$, as shown in Fig.~\ref{fig2}c. 

Before concluding this section, we note that a reliable numerical evaluation of the
above configuration averages -- as the one presented in
Fig.~\ref{fig2} -- is considerably more difficult to obtain than in
standard percolation problems, where the probability distribution of
the particles position is exactly known and systems configurations can
be generated from it. In our case, on the contrary, one has
to generate sufficiently uncorrelated configurations from the dynamics.  
This requires first to evolve the system from some initial condition into the stationary state (which for
large systems may require a considerable number of timesteps).
Then, in order to obtain configuration averages, one has to take averages over timescales $T$
much larger than the typical autocorrelation time. Consider for instance the 
timeseries of $s_M$ and $\ell_M$ discussed above (an
example of which for $\rho=1.5$ is shown in the inset of Fig.~\ref{fig2}d).
For the above parameters $L=256$ and $\eta=0.2$, the typical
autocorrelation time $\tau$ \cite{note3} in the low density regime are of the order of $10^4$ timesteps.
Note however that near the percolation transition
$\tau$ drops suddenly by almost two orders of magnitude. This could
seem counterintuitive, as phase transitions are typically associated
with a slowing down of the dynamics. However, one has to realize that 
near the percolation transition one has a wide distribution of
competing clusters with sizes close to the spanning threshold, so that relatively small
configuration changes may promote a different cluster to the largest
cluster status (either in total mass or linear extension) thus resulting in a dramatic drop
in the autocorrelation time for $s_M$ and $\ell_M$.

Once the autocorrelation time has been estimated, the accuracy of the empirical
averages can be evaluated by the standard error $\sigma /
\sqrt{T/\tau}$, where $\sigma$ is one standard deviation and $T/\tau$
the number of independent configurations.

\section{Finite size scaling analysis of percolation in the Toner-Tu liquid phase}
\label{sec:FSS}

One of the best ways to numerically investigate critical phase
transitions is to perform a finite-size-scaling (FSS) study, measuring the 
lowest moments of suitable order parameters as the system size is
systematically increased. This is a classical approach in statistical
physics, routinely applied to study both equilibrium and out-of-equilibrium
critical phase transitions \cite{FSS}, and it has been already
applied to the study of the percolation transition, for instance bond
percolation on square 
lattices \cite{Stauffer80, Stauffer}. The main difficulty, generally, is to be sure to probe system sizes 
large-enough so that one is in the scaling regime.

\subsection{Percolation in the longitudinal direction}

We first concentrate on the longitudinal percolation transition,
i.e. the point at which the spanning cluster becomes two-dimensional
and starts to span also in the broken symmetry direction.

The mean largest cluster size $n$
measures the probability $n/N$ that
an arbitrary particle belongs to the largest cluster.
In percolation theory, it is known to follow the finite size scaling relation \cite{Stauffer80}
\be
n = L^{-\beta/\nu} f((\rho-\rho_c^\infty)L^{1/\nu})\,,
\label{scaling1}
\ee
where $f$ is a scaling function and $\beta$ and $\nu$ two universal
critical exponents. At $\rho=\rho_c^\infty$, the asymptotic critical
point, $f(0)=\mbox{const.}$ and one obtains the power-law behavior $n \sim L^{-\beta/\nu}$. 
In two dimensional standard percolation, one has $\nu_{\rm p}=4/3$
and $\beta_{\rm p}=5/36$ \cite{Saberi} (and thus $\beta_{\rm p} / \nu_{\rm p} =
5/48$).

\begin{figure}[t!]
 \includegraphics[width=\columnwidth]{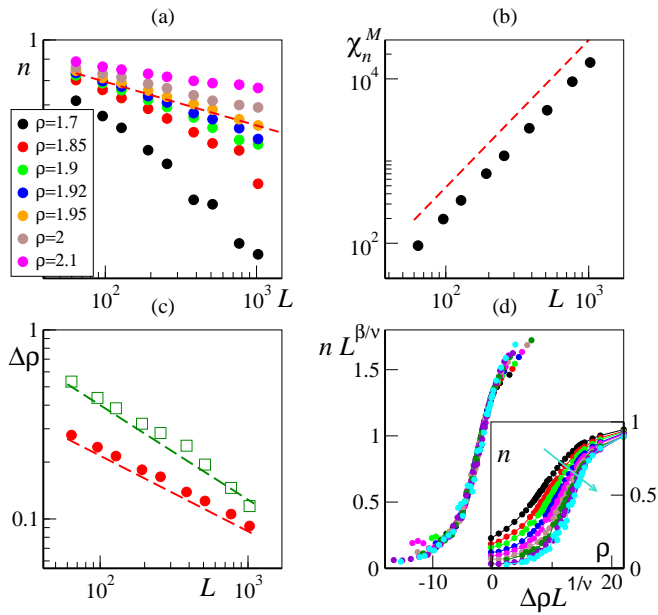} 
 \caption{Finite size scaling analysis of the percolation transition in the Toner-Tu ordered phase.
  (a) mean largest cluster size $n$ vs. system size $L$ for different densities
    (see legends). The dashed red line marks the standard
   percolation critical exponent ratio $\beta_{\rm p} / \nu_{\rm p} =5/48$. 
   (b) susceptibility peak value (black dots) $\chi_n^M$ vs. system size $L$. The dashed red line marks the standard
   percolation critical exponent ratio $\gamma_{\rm p} / \nu_{\rm p} =43/24$. 
(c) Critical point location finite size corrections $\Delta \rho=\rho_c^\infty
- \rho_c(L)$ evaluated either from the midpoint of the spanning
probability $S_2$ (red full dots) or from the
peak location of the standard deviation $\sigma_n (\rho)$ (green empty squares). 
Here we have used $\rho_c^\infty = 1.96$. The dashed red line marks a power law decay with an exponent $-0.4$, while
the dashed green line falls off as $L^{-0.5}$.
(d) Data collapse of $n$ according to the scaling relation (\ref{scaling1})
with $\rho_c^\infty=1.96$ and exponents $\beta=\beta_{\rm p}=5/36$, $1/\nu_\parallel=0.4$ for different system sizes
between $L=64$ and $L=1024$. Inset: Non collapsed curves. From top to
bottom: $L=64$, $L=96$, $L=128$, $L=192$, $L=256$, $L=384$, $L=512$,
$L=768$ and $L=1024$.
Other parameters are $\eta=0.2$ and $v_0=0.5$.
 As in Fig.~\ref{fig2},
   averages have been computed sampling a $10^6$ timesteps long timeseries 
    every $100$ timesteps. The standard error (see text) of the data shown in
   panels (a)-(c) is equal or smaller than symbol size.}
\label{fig3}
 \end{figure}

By systematically changing the
density $\rho$ and the system size between $L=64$ and $L=1024$, we 
find (see Fig.~\ref{fig3}a) that for $\rho_c^\infty \approx 1.95$ the 
mean largest cluster size indeed follows a power-law decay with an
exponent compatible (our best fit being $ \beta / \nu = 0.108(5)$) 
with the standard percolation value of $\beta_{\rm  p} / \nu_{\rm p} = 5/48$.

Another independent exponent can be deduced from the finite size
scaling of the maximum of the susceptibility $\chi_n \equiv L^2 \sigma_n^2$,
\be
\chi_n^M = L^{-\gamma/\nu} \,,
\ee
with -- for $d=2$ standard percolation -- $\gamma_{\rm p}=43/18$
\cite{Saberi} so that $\gamma_{\rm p} / \nu_{\rm p} = 43/24$. For each system size
$L$ we estimate the peak susceptibility $\chi_M$ by a
quadratic fit of the peak region of $\sigma_n(\rho)$. In Fig.~\ref{fig3}b
we show that once again our numerical estimates are in very good
agreement with the standard percolation exponent. Indeed, our best fit
of $\gamma_\parallel / \nu_\parallel = 1.83(5)$ is fully compatible with the
standard percolation value of $43/24$.

We are left with the estimation of the correlation exponent $\nu$ that
determines finite size corrections to the critical point,
\be\Delta \rho \equiv \rho_c^\infty - \rho_c(L) \sim L^{-1/\nu}\,.
\ee 
Here we adopt and compare two different estimates for the finite size
critical density $\rho_c(L)$. We first estimate it as the location $\rho_M$ of
the maximum of the largest cluster size standard deviation
$\sigma_n(\rho)$ (once again, evaluated through a quadratic fit of the
peak region). Our results, illustrated in Fig.~\ref{fig3}c (green
squares) essentially confirm our previous estimate for the asymptotic critical point,
 $\rho_c^\infty=1.96(1)$. However, finite size corrections decay slower
 than what expected for standard percolation in two dimensions, and we
 have $1/\nu_\parallel \approx 0.5$. 
 It has to be noted that this
 estimate is based on a second moment (the standard deviation), so
 that its reliability could be questioned.

A second, and perhaps more accurate estimate of the finite size
critical density can be obtained measuring the 
density value by which the finite size spanning
probability crosses $1/2$. We are here interested in the longitudinal
spanning probability $S_\parallel$. Measuring it in relatively small
systems, where the mean orientation ${\bf V}(t)$ strongly diffuses in
its anngular component, can be however a difficult task. The central limit theorem implies that
the mean orientation should diffuse with an angular diffusion constant
proportional to $\eta^2/N$. For small enough system sizes, thus, the
mean orientation can change faster than the time needed by clusters to
re-align transversally w.r.t. ${\bf V}(t)$. Therefore, in our FSS
analysis we find preferable to consider, instead of the transversal and longitudinal spanning
probabilities the one and two dimensional spanning probabilities $S_1$
and $S_2$. The former is the probability that a cluster spanning
along at least one spatial direction (i.e. to join two opposite sides
of the system) does exist. The latter probability,
on the other hand, requires the spanning cluster to wrap along both
spatial directions, that is to join all four sides of our
system. Numerical simulations show that -- at least in the parameter
range we are interested in --  for $L \gtrapprox 256$ we have to a
good accuracy $S_\perp \approx S_1$ and $S_\parallel \approx
S_2$. For smaller system sizes, however, we have $S_1 < S_\perp
< S_\parallel <S_2$. 

In the following, we estimate the finite size
critical density as the density value by which an error function based fit
\cite{note-err} of the finite size spanning probability $S_2(\rho)$
crosses $1/2$ (see Fig.~\ref{fig4}a).
This second estimate, reported by full red circles in
Fig.~\ref{fig3}c, also points towards $\rho_c^\infty=1.96(1)$, but with
an even slower decay of finite size corrections, $1/\nu_\parallel \approx 0.4$.

Altogether, our estimates for the critical
 exponent $\nu_\parallel$ are clearly different from standard percolation in
 $d=2$, $\nu_{\rm p}=4/3$. Combining our two different approaches we get
 $\nu_\parallel=2.2(3)$, with the upper limit $\nu_\parallel\approx 2.5$ being suggested by the
 slightly more reliable spanning probability estimates. An estimate
 exclusively based on the latter estimate would return $1/\nu_\parallel=0.40(2)$
 and $\nu_\parallel = 2.5(1)$.

This value for the critical exponent $\nu_\parallel$, different from
the one of standard percolation, is indeed confirmed by attempting a data collapse
of the mean largest cluster size $n$ according to the scaling relation
(\ref{scaling1}). Our data clearly rule out the value $\nu_{\rm p}=4/3$, and
we obtain a satisfactory collapse with $\beta_\parallel/\nu_\parallel=\beta_{\rm p}/\nu_{\rm p}=5/48$ and
$1/\nu_\parallel$ in the range $0.4 \sim 0.5$.

Also note that our asymptotic critical density $\rho_c^\infty=1.96(1)$ is significantly larger than the
 asymptotic critical density for standard continuum percolation: In two spatial dimensions, the most accurate
estimate for the continuum percolation threshold for non interacting,
fully penetrable disks of radius
$r$ randomly distributed according to a Poisson Point Process (PPP) corresponds to a critical area $A_c=\pi r^2 \rho_c^{PPP}=1.2808737(6)$
\cite{CPerc2}. Since a unit interaction radius corresponds to
a disk radius $r=1/2$, we have $\rho_c^{PPP}=1.43632545(9)$.

\subsection{Harris criterion for percolation in correlated density fields}

While a shift in the critical percolation point is not surprising in
the presence of activity, and indeed has been observed
before in disordered active matter systems \cite{Berthier2014}, the
significant difference between our estimate for the critical exponent $\nu$
and the standard percolation value $\nu_{\rm p}$ deserves a few more
comments.

It is indeed known that long-range
correlations in the particle density can change the value of the
critical exponent $\nu$. In the percolation literature, this is known
as the Harris criterion \cite{Harris, Weinrib}: in the presence
of sufficiently long-ranged density correlations 
\be
C_{\rho} (r) \sim r^{-\alpha} \;\; {\rm with} \;\;\; 
\alpha < \frac{2}{\nu_{\rm p}} \,,
\ee
finite size corrections are indeed stronger and
the exponent $\nu$ takes larger values:
\be
\nu=\nu_H = \frac{2}{\alpha} \,.
\ee
On the other hand, for correlations decaying faster, $\alpha > 2/\nu_{\rm p}$,
correlations are not relevant and usual finite size corrections apply,
$\nu=\nu_{\rm p}$.

Applying the Harris criterion to our results suggests that the density field correlation should
decay with a power-law with an exponent $\alpha=2/\nu_\parallel$ in the range $0.8
\sim 1$. Using only the estimate derived from the spanning probability distributions, we would
have $\alpha=0.80(4)$.

We recall that the Toner-Tu phase is endowed
with long-range density correlations \cite{TT1,
  TT2}. Their exact real space expression, however, is
not known explicitly, so that here we
resort to estimate them numerically in the range of sizes accessible to
the present FSS analysis. While it is known that correlations are stronger in the transversal
than in the longitudinal direction, the numerical measure of anisotropic correlations
is a challenging issue. Restricting the measure either in the
transversal or longitudinal directions greatly reduces the available
statistics and suffers from problems due to the angular diffusion of
the mean direction of motion analogous to the one discussed in the
previous section. On the other hand, one can expect that the onset of
a cluster percolating in both directions (as measured by the spanning probability $S_2$) could be well captured by
measures of density correlations averaged over all spatial
directions. In the following, therefore, we focus on 
isotropic correlations of density fluctuations
\be
C_\rho(r, L) = \langle \langle \delta \rho ({\bf x} + {\bf r}, t)\,\delta\rho({\bf
  x},t)\rangle_S \rangle_t
\label{corr}
\ee
where $r=|{\bf r}|$ and $\delta \rho ({\bf  x},t) \equiv \tilde{\rho}({\bf x}) - \rho$
are the local density fluctuations of a suitably coarse-grained
density field $\tilde{\rho}({\bf x})$, and $\langle \cdot \rangle_S$
indicates an average over the spatial coordinate ${\bf x}$ and the
orientations of the displacement ${\bf r}$. Isotropic correlations are
then further averaged in time, with $\langle \cdot \rangle_t$ indicating an
average over stationary state configurations.
A more detailed analysis of anisotropic correlations will be reported
in \cite{Benoit}.

\begin{figure}[t!]
 \includegraphics[width=\columnwidth]{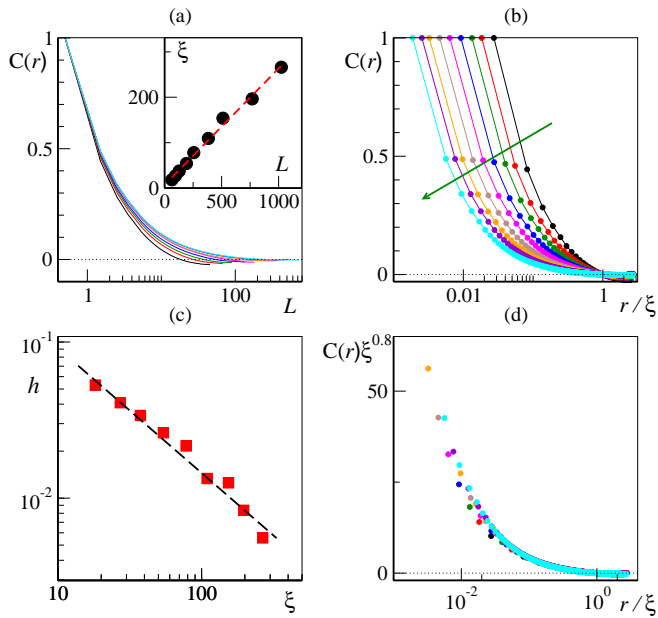} 
 \caption{(a) Isotropic density fluctuations correlation function in the Toner-Tu ordered phase  
   ($v_0=0.5$, $\eta=0.2$, $\rho=1.9$) and increasing system sizes,
   $L=64,96,128,192,256,384,512,768,1024$ (from bottom to top). 
   Inset: Correlation length $\xi$ as a function of system size $L$. The dashed red line marks our
   best linear fit.
(b) Same as (a) but after rescaling of the space variable. System size increases along the green arrow. 
(c) Finite size scaling of the (negative) slope $h$ (see text). The dashed
black line marks a power-law decay with exponent $-\alpha=-0.8$.
(d) Data collapse according to Eq. (\ref{CR}). 
Correlation functions have been averaged over $10^4$ different spatial
configurations, sampled from the stationary state dynamics every $10^2$
timesteps. Standard errors are of the size of the symbols or
smaller.}
\label{fig5}
 \end{figure}

The scaling of correlations is expected to be the same in the entire Toner-Tu phase.
Here we focus on a point close to the percolation threshold, $\eta=0.2$ and $\rho=1.9$,
but we have verified that the behavior for lower or higher densities stays the same.
Our numerically determined correlations are shown in
Fig.~\ref{fig5}a. Note that in finite systems the spatially integrated
fluctuations vanish by construction $\int d{\bf x} \delta \rho({\bf
  x},t)$, and this implies that the correlation function $C(r,L)$ should have at least
one zero (as there are surely anti-correlated regions). The smallest
value of $r$ for which correlations vanish 
can be taken as a measure of the correlation length $\xi$, that is
$C(\xi,L)=0$. 
Moreover, in systems
where a continuous symmetry is spontaneously broken correlations are known
to be scale free, i.e. $\xi \sim L$ (see inset of  Fig.~\ref{fig5}a) and $C(r) \sim r^{-\alpha}$ in the
thermodynamic $L \to \infty$ limit.
Finite size correlations are thus taken into account by \cite{CavagnaScaleFree}
\be
C_\rho(r,L) = r^{-\alpha} \,g\left(\frac{r}{\xi}\right) \,,
\label{fs}
\ee
where the scaling function obeys $g(u) = 0$ for $u=1$ and  $g(u) \to \mbox{const.}$ for $u \to 0$.

The isotropic correlation exponent $\alpha$ can be determined by
finite size analysis. Let us choose the rescaling $y=r/\xi$. From
Eq. (\ref{fs}) we have (see Fig.~\ref{fig5}b)
\be
C_\rho(y,L) = y^{-\alpha} \xi^{-\alpha} g(y) \,.
\label{fsr}
\ee
From Eq. (\ref{fsr}) it follows that a finite size analysis of the (negative) slope $h$ of the rescaled correlation function evaluated
in $y=1$ can be used to estimate the correlation exponent,
\be
h = -\frac{d}{dy}  \left.  C_\rho(y,L)\right|_{y=1} = \xi^{-\alpha} \left|
g'(1) \right| \sim \xi^{-\alpha} \,.
\label{fss-a}
\ee
Our best numerical estimates, reported in Fig.~\ref{fig5}c, are indeed
compatible with the correlation value suggested by the Harris criterion,
$\alpha \approx 0.8$. 

Moreover, as illustrated in Fig.~\ref{fig5}d, once the correlation exponent has been determined, the
finite size correlation functions can be collapsed to a size
independent universal curve
\be
C_\rho^R(y) \equiv \xi^{-\alpha} C_\rho\left(\frac{r}{\xi},L\right) \,.
\label{CR}
\ee

Our brief analysis of density correlations shows that the anomalous
finite size corrections exponent $\nu$ we have measured for our
percolation transition (especially through the more reliable spanning
probabilities measures) is fully compatible with the one expected by
the Harris criterion for correlated percolation.

We finally note that also the critical exponents $\beta$ and $\gamma$ may be
modified by sufficiently strong correlations. However, it has
been however verified numerically \cite{Saberi} that the hyperscaling
relation of standard percolation 
\be
\nu_{\rm p} d_s = 2 \beta_{\rm p} + \gamma_{\rm p}
\label{hypers}
\ee
(with $d_s$ being the spatial dimension) is still verified by the
correlated Harris exponents
\be
\nu_H d_s = 2 \beta_H + \gamma_H \,.
\ee
Interestigly, we find that this latter hyperscaling relation is also verified
by our data: as we have seen, our two dimensional estimates for
the ratios $\beta/\nu$ and $\gamma/\nu$ are both compatible with the values
expected by standard correlation theory, so that
\be
 2 \frac{\beta}{\nu} + \frac{\gamma}{\nu} = 2.05(6) \,.
\ee

\subsection{Percolation in the transversal direction}

\begin{figure}[t!]
 \includegraphics[width=\columnwidth]{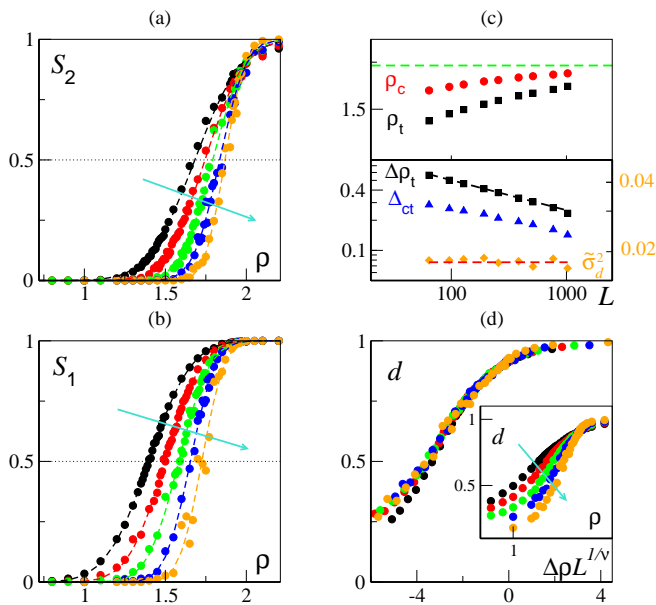} 
 \caption{(a)-(b) Two- and one-dimensional spanning probabilities $S_2$
and $S_1$ as a function of global density $\rho$ for different
system sizes ($L=64,128,256,512,1024$, increasing along the cyan arrow.
Dashed lines are fits by the error function \cite{note-err},
while the horizontal dotted line shows the threshold probability $1/2$
used to define the finite-size percolation point (see text).
(c) Top panel: Transversal (black squares) and longitudinal (red
circles) finite-size percolation densities as a function of system size $L$. 
The horizontal green line marks our best estimate for the asymptotic
percolation point $\rho_c^\infty = 1.96$. (c) Bottom panel: 
Transversal percolation point finite-size corrections $\Delta \rho_t=\rho_c^\infty
- \rho_c(L)$ (black squares), longitudinal to transversal finite
size difference $\Delta_{c\,t}$ (blue triangles, see text) and maximum variance of the largest cluster linear
extension $\tilde{\sigma}_d^2$ as a fuction of system size $L$ in a double
logarithmic scale. The dashed black
line marks a power-law decay with an exponent $0.3$, while the red one
corresponds to constant behavior.
(d) Data collapse of the mean maximum cluster extension $d$ according to the scaling relation (\ref{scaling2})
with $\rho_c^\infty=1.96$ and $1/\nu_\perp=0.3$ for different system
sizes between $L=64$ and $L=1024$  (color coded as in panels (a) and
(b)). In the inset: Non collapsed curves for
the mean largest cluster size $L$ vs. density $\rho$. Along the cyan arrow: $L=64, 128, 256, 512, 1024$. 
System parameters and simulation
statistics as in Fig.~\ref{fig3}.
}
\label{fig4}
 \end{figure}

We finally discuss the percolation transition taking place in the
transversal direction. Repeating the procedure outlined for the
longitudinal percolation transition, we evaluate the finite size
transversal percolation threshold density $\rho_{\rm t}(L)$
from the behavior of the one dimensional spanning probability $S_1$
(see Fig.~\ref{fig4}b). Our best estimates are reported in the top
panel of Fig.~\ref{fig4}c (black squares) together with the ones for
the finite-size longitudinal percolation critical density $\rho_c(L)$ (red
circles). Quite interestingly, our numerical results indicate that
their difference $\Delta_{c\,t}(L) \equiv \rho_c(L)-\rho_t(L)$
(blue triangles in the lower half of Fig.~\ref{fig4}c)
seems indeed to vanish in the limit of large $L$, suggesting that in
the thermodynamic limit $\rho_t(L) \to \rho_c^\infty=1.96(1)$ and 
percolation takes place simultaneously in both directions.

Finite size effects, however seem to be stronger in the transversal direction, with a slower decay
of finite size corrections 
\be
\Delta \rho_t (L) \equiv \rho_t(L) - \rho_c^\infty \sim
L^{-1/\nu_\parallel} \,.
\ee
This can be deduced from the lower panel of Fig.~\ref{fig4}c), where 
$\Delta \rho_t (L)$  (black squares) exhibits a power-law decay
compatible with an exponent $1/\nu_\perp \approx 0.3$, suggesting
$\nu_\perp \approx 3.3$.

Going beyond transversal finite size effects, however, we notice that
the cluster maximum linear extension $d$ seems to show rather
anomalous scaling properties. As it can be readily deduced from the inset of 
Fig.~\ref{fig4}d, its finite size curves do cross near the critical
density. This implies that its scaling relation should take the form
\be
d = f_\perp((\rho-\rho_c^\infty)L^{1/\nu_\perp})
\label{scaling2}
\ee
with $f_\perp$ a transversal scaling function. As we show in Fig.~\ref{fig4}d, one can indeed make
use of this scaling relation to achieve a satisfactory collapse of the
$d(\rho)$ curves by only rescaling them along the abscissas. 
Comparison with the general scaling form (\ref{scaling1}) thus implies the rather singular $\beta_\perp=0$.

Finally we discuss the $\gamma$ exponent, associated to the maximum linear extension susceptibility 
\be
\chi_d \equiv L^2 \sigma_d^2,
\label{chid}
\ee
whose peak value is expected to scale as 
\be
\chi_d^M \sim L^{\gamma_\perp/\nu_\perp}\,.
\ee
Assuming that an hyperscaling relation analogous to (\ref{hypers})
still holds between the transversal exponents, in two spatial dimensions
we would get $\gamma_\perp/\nu_\perp \approx 2$. By virtue of Eq.
(\ref{chid}), this in turn implies that also the peak variance of
the largest cluster linear extension should not scale with system
size, 
\be
\tilde{\sigma}_d^2 (L) \equiv \mbox{max}_\rho \sigma_d^2 (\rho, L)
\sim \mbox{const.}
\ee
This is indeed verified by our
numerical data (see the bottom panel of Fig.~\ref{fig4}c, where the
maximum has been evaluated by a quadratic fit of the peak region). 
We conclude that the cluster maximum linear
extension $d$ shows no finite size scaling, apart from finite
size corrections in its density dependence.

\subsection{Anisotropic percolation exponents}

Our estimates for the Toner-Tu phase
percolation exponents in two spatial dimensions, as measured from simulations of the Vicsek model,
are summarized in Table~\ref{table1}
 
\begin{table}[h!]
\caption{Longitudinal and transversal percolation exponents compared
  with the ones of standard percolation theory. Longitudinal exponents
  are particularly difficult to evaluate due to strong finite size
  effects and ours are rough estimates. For this reason we are not
  confident in providing precise uncertainty estimates.}
\begin{ruledtabular}
\begin{tabular}{lccc}
 & $1/\nu$ & $\beta/\nu$ & $\gamma/\nu$ \\ 
 Standard percolation, $d_s=2$ & 3/4  & 5/48 & 43/24 \\ 
 Longitudinal percolation & 0.44(6) & 0.108(5) & 1.83(5)\\  
 Transversal percolation & 0.3 & 0 & 2   \\
\end{tabular}
\end{ruledtabular}
\label{table1}
\end{table}

In the Toner-Tu theory, anisotropy is controlled by the exponent $\xi$
\cite{TT1}, so that one should expect $\nu_\perp = \nu_\parallel/\xi$ or
\be
\xi = \frac{\nu_\parallel}{\nu_\perp} \approx 0.6 \sim 0.75 \,.
\ee
In two spatial dimensions, based on some
renormalization group conjectures, Toner and Tu suggested \cite{TT2,
  TT3} that $\xi = 3/5$, a value which coincides with the lower bound
of our FSS measure. However, it should be noted that the more reliable spanning
estimates rather support the upper bound of our estimates $\xi \approx
0.75$, thus suggesting a less severe anisotropy. More precise measures
will be required to shed light on this issue \cite{Benoit}.

Before concluding this section, we briefly comment on the behavior of
the percolation threshold as a function of the Vicsek noise amplitude
$\eta$. While a careful determination of the full percolation line is
beyond the scope of this work, preliminary simulations indicates that
for $\eta=0.1$ one has $\rho_c^\infty = 2.2(1)$ (see Fig.~\ref{PD}), suggesting that the
percolation critical density in the TT phase should be a decreasing
function of noise amplitude.

\section{Cluster size distribution in the Toner-Tu liquid phase}
\label{sec:CSD}

We proceed to discuss cluster size distributions, a widely used quantity both
in percolation theory and in the literature on non-equilibrium clustering in active systems.

The cluster size distribution (CSD) is one of the simplest objects to be
computed numerically in percolation theory. 
One should notice though, that CSD corresponds to two different meanings in the literature.
In a first approach, the CSD $P(s)$ measures the (properly normalized)
number of clusters with size $s$ one finds in given configurations. This corresponds in
practice to the probability to find a cluster of size $s$ when we pick
at random one of the many clusters we identify in our dynamics.
Other authors, however, prefer to work with the probability $Q(s)$
that a particle picked at random belongs to a cluster of size
$s$. Obviously the two measures are related, $Q(s) = s P(s)$, so
that the choice between the two above definitions is equivalent. In the
following we consider $P(s)$. We measure it by sampling
a large number (typically $10^4$) of different steady-state configurations 
of our dynamics, obtained from a single run (after a dynamical
transient has been discarded), with 100 time units separating
consecutive configurations.
In the following, we may also find convenient to further rescale the cluster size
$s$ by the total number of particles $N$, so that we deal with a
normalized cluster size variable $s/N \leq 1$.

\begin{figure}[t!]
 \includegraphics[width=\columnwidth]{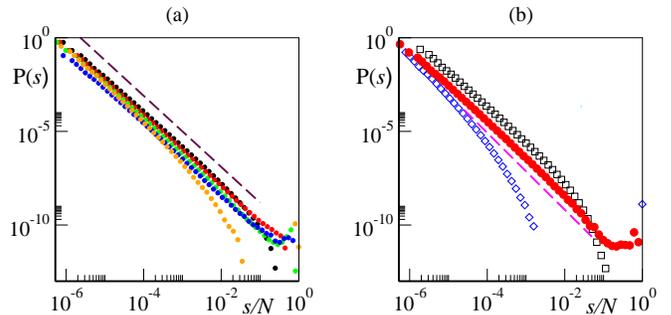} 
 \caption{(a) Cluster size distribution P(s) vs. the rescaled cluster
   size $s/N$ at different densities in the Toner-Tu ordered phase (from top to bottom: 
   $\rho=1.3$ (black), $\rho=1.6$ (red), $\rho=1.9$ (green),
   $\rho=2.2$ (blue), $\rho=2.5$ (orange). The dashed line marks marks
   a power-law decay with a exponent equal to $-1.9$.
(b) $P(s)$ at the percolation threshold ($\rho=1.95$, full red circles), compared with
off-critical values $\rho=1$ (black squares) and $\rho=4$ (blue
diamonds) for  $L=1024$.
The magenta dashed line marks
a power-law decay with the Fisher exponent $\tau_F=187/91\simeq 2.0549$.
All distributions are log-binned, and have been computed sampling a $10^6$ timesteps trajectory every $100$ timesteps. 
Other parameters: $L=1024$,  $\eta=0.2$ and $v_0=0.5$.}
\label{fig6}
 \end{figure}

We have measured the CSD in the ordered liquid phase  along the dashed blue line in the phase diagram
of Fig.~\ref{PD}, that is, at noise amplitude $\eta=0.2$. Our results, reported in
Fig.~\ref{fig6}a, suggest that cluster size in the desity interval
$\rho \in [1.2,2.2]$ follows a power-like behavior over a wide range of
scales (about four decades for the size considered). 
This is in agreement with previous studies \cite{Ginelli2008}.

Considering density values further out from the
percolation point, see for instance Fig.~\ref{fig6}b, 
one observes clear exponential cut-offs from power-law behavior.
Note that above the percolation density, $\rho >\rho_c^\infty$, where a single giant
connected cluster typically appears, the CSD shows an
exponential cut-off, but also, beyond that, a finite probability of observing clusters
of size $s \approx N$.

\begin{figure}[t!]
 \includegraphics[width=\columnwidth]{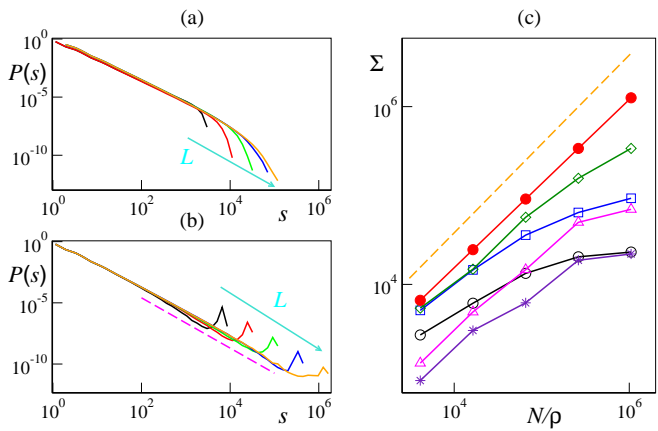} 
 \caption{
   (a) Finite size variation of $P(s)$ for $\rho=1$ 
   (following the cyan arrow $L=64,128,256,512,1024$).
(b) same as (b), but at the percolation threshold $\rho=1.95$, where
no size-dependent cut-off is present. The magenta dashed line marks a
power-law decay with the Fisher exponent $\tau_F=187/91$.
(c) Estimated cut-off length $\lambda$  (see text) as a function of
$N/\rho$ for different density values: $\rho=1$ (black circles),
$\rho=1.3$ (blue squares), $\rho=1.6$ (green diamonds),
$\rho=1.95$ (full red circles), $\rho=2.2$ (magenta triagles),
$\rho=2.5$ (indigo stars). The dashed orange line marks the linear
relation $\sim N$.
All distributions are log-binned, and have been computed sampling a
$10^6$ timesteps trajectory every $100$ timesteps. Other parameters:
$\eta=0.2$ and $v_0=0.5$. All panels are in a double logarithmic scale.}
\label{fig6bis}
 \end{figure}

The proper way to discriminate a true power-law behavior
from an approximate one is, once again, finite-size analysis. We considered 
systems of different sizes between $L=64$ and $L=1024$. Off-critical CSDs, as the
ones shown in Fig.~\ref{fig6bis}a, exhibit an exponential cut-off at
size $\Sigma$. While $\Sigma$ may initially grow
with system size, finite size analysis of its estimated
value \cite{noteSigma} shows saturation effects towards an asymptotic
value $\Sigma^\infty(\rho)$. As shown in
Fig.~\ref{fig6bis}c, this saturation seems to occur for
all densitiy values different from the critical percolation density,
with $\Sigma^\infty(\rho)$ increasing as the percolation threshold is
approached from both sides.
This implies that the power laws reported in Fig.~\ref{fig6}a are not asymptotic.
It is only at the anisotropic percolation point $\rho \approx \rho_c^\infty=1.96(1)$ discussed in the
previous sections that $\Sigma^\infty(\rho)$ diverges, and a truly asymptotic critical CSD appears. 
CSDs at the percolation threshold at different system sizes are reported in
Fig.~\ref{fig6bis}b. They show a large size peak corresponding to the
typical size of the percolating cluster, which is clearly scaling with
the system size $N$, as it can also be appreciated from
Fig.~\ref{fig6bis}b, where we have used the location of this peak to
estimate the critical point typical cluster size $\Sigma$ (full red dots).

We finally estimate the power law decay exponent at the percolation point.
At the critical point of standard percolation, the cluster size
distribution power law behavior is controlled by the so-called
Fisher exponent,
\be
\tau_F=\frac{2d_s-\beta/\nu}{d_s-\beta/\nu}\,,
\ee
which only depends on the spatial dimension $d_s$ and on the critical
exponents ratio $\beta/\nu$ \cite{Stauffer}.
In two spatial dimensions we get $\tau_F=\frac{187}{91}\approx 2.05$. We
have seen that in the longitudinal percolation transition, the scaling
of the largest cluster size $n$ is still controlled by the standard
percolation exponent ratio $\beta_{\rm p}/\nu_{\rm p}$, so that we also expect
our cluster size distribution near the critical percolation
density $\rho_c^\infty \approx 1.96$ to behave as in standard
percolation, that is
\be
P(s) \sim s^{-\tau_F} \,.
\ee

This is indeed verified by our data. For $\rho=1.95$, the
corresponding CSD (full red dots in Fig.~\ref{fig6}b) exhibits a power law behavior fully
compatible with the standard Fisher exponent (orange dashed line) over
several decades. Our best fit, carried on over roughly four decades,
gives indeed $\tau = 2.03(3)$. Note that this value is different from that of
the apparent, non asymptotic power laws observed at $\rho \neq \rho_c^\infty$, which
have been found typically in the range $[1.8, 2]$ \cite{Ginelli2008, Huepe}.  

Altogether, our results show that, while truly critical CSDs only appear at the 
percolation point, the Toner-Tu ordered phase nevertheless displays an
extended ``quasi-critical'' region, where cluster size distributions
follow a power-law over several orders of magnitudes and 
for a wide range of densities. 
This approximate critical regime has also been reported in previous
works \cite{Ginelli2008, Huepe,Peruani2013}
and -- as we have discussed in the introduction -- has led some authors to speculate that the onset of
collective motion should be accompanied by a
percolation transition. 
The analysis of the anisotropic percolation
transition carried on in the previous chapter however, clarifies that
the Toner-Tu phase of finite size systems is characterized by a ``double''
percolation transition, with giant clusters first percolating
transversally w.r.t. the mean direction of motion and, at higher
densities, also spanning in the longitudinal direction.
We conjecture that this extended region of scaling is related
to the two separate finite size transitions
at two clearly different densities. Note also that far away from
this ``extended region'', the cluster size distributions are clearly
not scale free, see for instance the case $\rho=4$, $\eta=0.2$ (blue
squares) in Fig. ~\ref{fig6}c.

\section{Percolation and clustering in the disordered and coexistence phases}
\label{sec:vcuts}

\begin{figure}[t!]
 \includegraphics[width=\columnwidth]{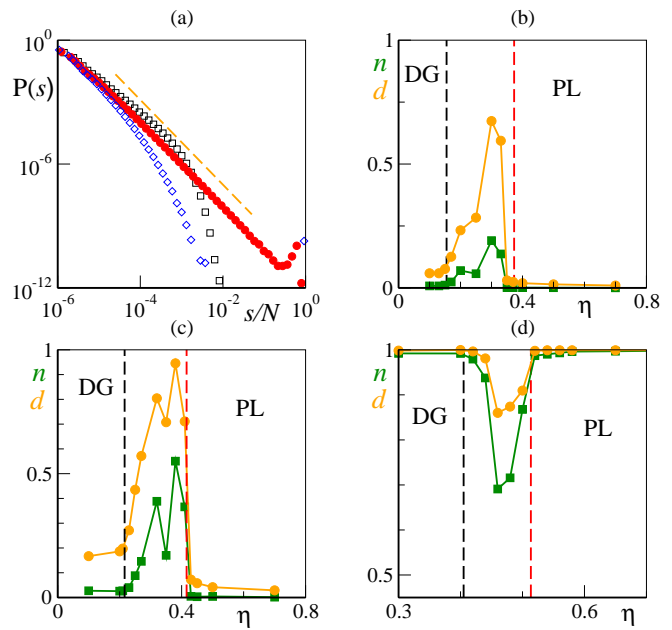} 
 \caption{(a) Cluster size distribution P(s) vs. rescaled cluster size
   $s/N$ for $\eta=0.7$ and different densities in the disordered gas phase($L=1024$); from top to bottom:
   $\rho=1.3$ (black squares),
   $\rho=1.51$ (the percolation point, full red circles), and
   $\rho=1.7$ (blue diamonds).
The orange dashed line marks
a power law decay with the Fisher exponent $\tau_F=187/91$.
(b,c,d) Mean value of the largest cluster size, $n$ (green squares) and maximum cluster
extension, $d$ (orange circles) vs. noise amplitude $\eta$ across all 3 phases. Parameters: 
$v_0 = 0.5$ and (b) $L = 1024$, $\rho=0.5$, (c) $L = 1024$, $\rho=1$ (d)
$L = 512$, $\rho=4$. The black vertical dashed lines mark the finite-size onset of
order \cite{noteL}, $\eta=\eta_{\rm gas}(L)$, while the red ones, at $\eta=\eta_{\rm liq}$
separate phase coexistence (between the two vertical lines)
from the Toner-Tu polar liquid (PL).} 
\label{fig7}
\end{figure}

A percolation transition is of course also found in the disordered gas
phase. It is a simple isotropic one with standard exponents. Its transition line is
reported in Fig.~\ref{PD}, and for maximal noise culminates at the
well known critical point for a Poisson Point Process,
$\rho_c^{PPP}=1.43632545(9)$, as discussed at the end of
Sec.~\ref{sec:FSS}. Note that also in this case, the short ranged
correlations arising in the disordered phase for noise amplitudes $\eta<1$
shift the critical percolation point to slightly larger
density values. Here, however, without the ``double'' finite size percolation
mechanism we have unearthed
in the symmetry broken regime, off-critical
cluster size distributions do not show any apparent power-law behavior
as their counterparts in the ordered liquid phase. See for instance
Fig.~\ref{fig7}a for noise amplitude $\eta=0.7$

In the coexistence phase delimited by the two binodal lines, where high-density high-order traveling bands
are observed, the cluster dynamics is radically different. 
We selected three different densities well below ($\rho=0.5$, $\rho=1$) and above ($\rho=4$) the
percolation transition lines of both the disordered gas and the Toner-Tu phases, and varied
the noise amplitude as shown in Fig.~\ref{PD}, in order to cut across both binodals. 
We computed both the largest cluster size, $n$ and the maximum cluster
extension, $d$. For low densities, data shows that in the gas and
Toner-Tu phases, clusters are small and do not reach a macroscopic, system
spanning state (Fig.~\ref{fig7}b-c). However, in the coexistence
region \cite{noteL}, high and low local density patches appear (signaling the presence of ordered
liquid bands travelling in a disordered gas) \cite{Solon2015}, and system spanning clusters suddenly
appear. On the other hand, at large densities (Fig.~\ref{fig7}d), in
both the gas and Toner-Tu phases, one has typically a single cluster
encompassing almost all particles, with $n,d \approx 1$. The appearance
of lower density disordered patches, on the other hand, induces a drop
in the maximum cluster size in the coexistence region. It has been
shown that due to these effects, also the cluster size distribution
built by averaging over both phases in the coexistence region show
apparent power laws, albeit with a decay exponent larger than the
Fisher one \cite{Ginelli2008}.

\section{Discussion and conclusions}
\label{sec:conclusion}

Our numerical results  show that nonequilibrium
clustering effects in the two dimensional Vicsek model are essentially controlled by an
underlying percolation point, and are therefore mainly geometrical in nature. 
Cluster dynamics and cluster
size distributions behave differently not only in the different
phases, but also within phases, as one always expects to cross a
percolation transition when the density is sufficiently large.  
Moreover, crossing one of the binodal lines delimiting the coexistence phase separating
the disordered gas form the Toner-Tu ordered liquid,
 sudden changes are typically observed in the cluster dynamics and 
 corresponding cluster size distributions. These transitions, however, are dictated by the
 overall phase-separation scenario of the phase diagram, and not vice-versa.

In the disordered gas phase, a standard percolation
transition is observed, akin to that observed at maximal noise (i.e. in a system
fully equivalent to a Poisson point process), with standard
percolation exponents \cite{Stauffer} but a slight shift
in the critical percolation density due to short range correlations.

In the Toner-Tu symmetry-broken phase, on the other hand, we have identified a novel
anisotropic percolation transition with clusters first
 spanning the transversal direction (w.r.t. the mean direction of
motion) and only later, at higher densities, spanning also along the
longitudinal direction. A careful finite size analysis revealed that
these two distinct percolation thresholds seem to converge to the same
density value in the thermodynamic limit, albeit with two different
correlation exponents $\nu_\perp$ and $\nu_\parallel$, which are
are also clearly different from the well-known
value of the standard percolation correlation exponent $\nu_{\rm p}$ in two
spatial dimensions. 

We have argued that the difference in the correlation exponents
can be attributed to the long-range correlations which characterize
 density fluctuations in the Toner-Tu phase. In particular, making
use of the Harris criterion \cite{Harris} for correlated percolation,
we have been able to link the value of the longitudinal correlation exponent 
(the one controlling the onset of a cluster of macroscopic mass
spanning in both directions) with the isotropic (i.e. averaged over
all directions) density fluctuation correlations.

The hyperscaling
relation of standard percolation seems to hold also in the
correlated Toner-Tu phase, with the key exponents controlling the cluster size
distribution (the Fisher exponent $\tau_F$) and the first two momenta
of the maximum cluster size ($\beta/\nu$ and $\gamma/\nu$) compatible
with their values from standard percolation theory.

In general, it is only at the percolation point that the cluster size
distribution is truly scale free ($P(s) \sim s^{-\tau_F}$). However,
cluster size distributions resembling power-laws over a wide range of scales
occur for a finite range of densities in the Toner-Tu phase, presumably
because of  the ``double-threshold'' mechanism of anisotropic percolation.
Only a careful finite size analysis can show that these power laws are not asymptotic
but bounded by a size-independent cut-off. 

Here we have closely analyzed clustering and percolation in the
classical Vicsek model for flocking, but we expect our main conclusions
to be generic and to hold in the more general context of dry aligning active matter.

\begin{acknowledgments}
We have benefited from discussions with F. Perez-Reche. NK and FG acknowledge support from the Marie
Curie Career Integration Grant (CIG) PCIG13-GA-2013-618399. FG also
acknowledges support from EPSRC First Grant EP/K018450/1.
\end{acknowledgments}


\begin{thebibliography}{99}
\bibitem{noteS}
Both the interaction range and the timestep can be rescaled to unit
values without any loss of generality.
\bibitem{note1} 
This is numerically much faster (order $s$ vs order $s^2$ in a cluster of $s$ particles) to compute than the maximum
  distance between any couple of particles in the cluster. Of course periodic boundary conditions
  are properly taken into account computing the cluster center of mass.
\bibitem{note3} 
We measure the autocorrelation time $\tau$ as the time needed for the
autocorrelation function to drop to $10\%$ of its equal time value.
\bibitem{note-err}
More precisely, we fit $S_2(\rho)$ by $F(\rho)\equiv\left[1 -
\mbox{erf}\left(a_1 \,(\rho-a_0) \right)\right]/2$, with the size-dependent
fitting parameters $a_0$ and $a_1$. Obviously, $\rho_c(L)=a_0$.
\bibitem{noteSigma}
In order to estimate the off-critical cut-off size $\Sigma$ from the
CSD, we first subtract the apparent power law decay. $\Sigma$
correspond to the cluster size at which this rescaled CSD drops below
the threshold $1/e$. We have verified that the choice of different
thresholds does not chage the qualitative picture.
\end{thebibliography}
\end{document}